\documentclass[aps,prl,numerical,linenumbers,superscriptaddress,showpacs,floatfix,reprint]{revtex4-1}
\usepackage[dvipdfm]{graphicx}
\usepackage{amsmath,amssymb,amsfonts}
\usepackage{color}

\newcommand{\be}{\begin{eqnarray}}
\newcommand{\ee}{\end{eqnarray}}

\newcommand{\sqrtsNN}{\mbox{$\sqrt{\mathrm{\it s_{NN}}}$} }

\def\v2{\mbox{$v_2$}}

\def\sqrtsNN{\mbox{$\sqrt{s_{NN}}$}}

\begin{document}


%
\title{Beam energy dependence of the viscous damping of anisotropic flow}
%
%
%
\author{ Roy~A.~Lacey}
\email[E-mail: ]{Roy.Lacey@Stonybrook.edu}
\affiliation{Department of Chemistry, 
Stony Brook University, \\
Stony Brook, NY, 11794-3400, USA}
\affiliation{Physics Department, Stony Brook University, \\
Stony Brook, NY, 11794-3800}
\author{A.~Taranenko}
\affiliation{Department of Chemistry, 
Stony Brook University, \\
Stony Brook, NY, 11794-3400, USA} 
\author{ J.~Jia}$^2$
\affiliation{Department of Chemistry, 
Stony Brook University, \\
Stony Brook, NY, 11794-3400, USA}
\affiliation{Physics Department, Bookhaven National Laboratory, \\
Upton, New York 11973-5000, USA}
\author{D. Reynolds} 
\affiliation{Department of Chemistry, 
Stony Brook University, \\
Stony Brook, NY, 11794-3400, USA}
\author{ N.~N.~Ajitanand} 
\affiliation{Department of Chemistry, 
Stony Brook University, \\
Stony Brook, NY, 11794-3400, USA}
\author{ J.~M.~Alexander}
\affiliation{Department of Chemistry, 
Stony Brook University, \\
Stony Brook, NY, 11794-3400, USA}
\author{Yi Gu} 
\affiliation{Department of Chemistry, 
Stony Brook University, \\
Stony Brook, NY, 11794-3400, USA}

\author{A. Mwai}
\affiliation{Department of Chemistry, 
Stony Brook University, \\
Stony Brook, NY, 11794-3400, USA}

%
%
%



\date{\today}


\begin{abstract}

	The flow harmonics $v_{2,3}$ for charged hadrons, are studied for a broad range of 
centrality selections and beam collision energies in Au+Au ($\sqrt{s_{NN}}= 7.7 - 200$ GeV) 
and Pb+Pb ($\sqrt{s_{NN}}= 2.76$ TeV) collisions. They validate the characteristic signature 
expected for the system size dependence of viscous damping at each collision energy studied. 
The extracted viscous coefficients, that encode the magnitude of the ratio of shear viscosity to 
entropy density $\eta/s$, are observed to decrease to an apparent minimum as the collision energy is 
increased from $\sqrt{s_{NN}}= 7.7$ to approximately 62.4~GeV; thereafter, they show a slow 
increase with $\sqrt{s_{NN}}$ up to 2.76 TeV. This pattern of viscous damping provides the first 
experimental constraint for $\eta/s$ in the temperature-baryon chemical potential ($T, \mu_B$) plane, and 
could be an initial indication for decay trajectories which lie close to the critical end point in the 
phase diagram for nuclear matter.  
	
\end{abstract}

\pacs{25.75.-q, 25.75.Dw, 25.75.Ld} 

\maketitle


	Heavy ion collisions provide an important avenue for studying the phase diagram for 
Quantum Chromodynamics (QCD) \cite{Itoh:1970,Shuryak:1983zb,Stephanov:1998dy}. The location of the phase boundaries 
and the critical end point (CEP), in the plane of temperature vs. baryon chemical 
potential ($T, \mu_B$), are fundamental characteristics of this phase diagram \cite{Asakawa:1989bq}. 
Lattice QCD calculations suggest that the quark-hadron transition is a crossover at high 
temperature ($T$) and small $\mu_{B}$ or high collision energy (\sqrtsNN)~\cite{Aoki:2006we}. 
For larger values of $\mu_{B}$ or lower \sqrtsNN~\cite{chemfo}, several model calculations have indicated a first 
order transition~\cite{Ejiri:2008xt,Stephanov:2005} and hence, the possible existence of a CEP. 
It remains an experimental challenge however, to validate many of the essential ``landmarks'' 
of the phase diagram, as well as to extract the properties of each QCD phase.

Anisotropic flow measurements are sensitive to initial conditions, the equation of state (EOS)
and the transport properties of the medium. 
Consequently, they are key to ongoing efforts to delineate the \sqrtsNN~or ($T, \mu_B$) 
dependence of the transport coefficient $\eta/s$, of the hot and dense matter created in collisions at both 
the Relativistic Heavy Ion Collider (RHIC) and the Large Hadron Collider (LHC).
The Fourier coefficients $v_n$ are frequently used to quantify anisotropic flow as a function 
of particle transverse momentum $p_T$, collision centrality (cent) and \sqrtsNN;
\begin{equation}
\frac{dN}{d\phi} \propto \left( 1 + 2\sum_{n=1}v_n\cos n(\phi - \psi_n) \right),
\label{eq:1}
\end{equation}  
where $\phi$ is the azimuthal angle of an emitted particle, 
and $\psi_{n}$ are the azimuths of the estimated participant event 
planes \cite{Ollitrault:1992bk,Adare:2010ux};
%
$
v_n = \left\langle \cos n(\phi - \psi_n) \right\rangle, 
$
%
%
where the brackets denote averaging over particles and events, for a given centrality and $p_T$ at 
each \sqrtsNN~\cite{yfluc}.
 
The LHC $v_n$ measurements at $\sqrt{s_{NN}}= 2.76$ TeV, allow investigations of $\eta/s$ at high $T$ 
and small $\mu_{B}$; they compliment the $v_n$ measurements from the recent RHIC beam-energy scan (BES) 
which facilitates a study of $\eta/s$  for the $\mu_B$ and $T$ values which span the collision energy 
range $\sqrt{s_{NN}}= 7.7 - 200$ GeV. Here, it is noteworthy that there are currently no experimental 
constraints for the $\mu_B$ and $T$ dependence of $\eta/s$, especially for the lower beam energies.
At the CEP or close to it, anomalies in the dynamic properties of the medium can drive abrupt  
changes in transport coefficients and relaxation rates \cite{Csernai:2006zz,Lacey:2007na}. Therefore, a study of 
$v_n$ measurements which span the the full range of energies available at RHIC and the LHC, also 
provides an opportunity to search for characteristics in the $\sqrt{s_{NN}}$ [or $(T, \mu_B)$] 
dependence of $\eta/s$ which could signal the location of the CEP \cite{Csernai:2006zz,Lacey:2007na}. 
 
	An important prerequisite for such studies is a method of analysis which allows a consistent 
evaluation of the influence of viscous damping on the $v_n$ measurements which span the full 
range of $\sqrt{s_{NN}}$ values. 
%
\begin{figure}[ht]
\includegraphics[width=1.0\linewidth]{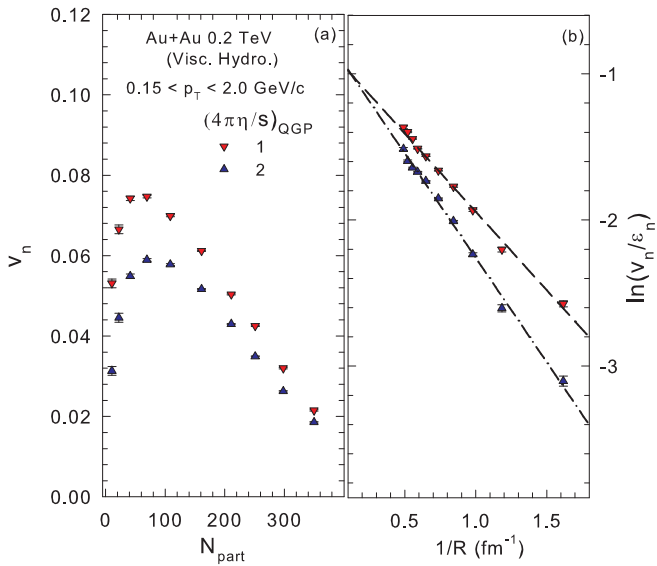}
\caption{(a) $v_{2}$ vs. N$_\text{part}$ from viscous hydrodynamical calculations \cite{Song:2011hk-a,Song:2011hk} for two 
values of specific shear viscosity as indicated. The results are for $0.15 < p_T < 2.0$ GeV/c for Au+Au collisions
at $\sqrt{s_{NN}}= 200$ GeV. (b) $\ln(v_n/\varepsilon_n)$ vs. $1/\bar{R}$ for the $v_2$ values shown in (a). 
The dashed and dot-dashed curves are linear fits.
}
\label{Fig1m}
\end{figure}
%
%
\begin{figure*}[t]
\begin{tabular}{cc}
\includegraphics[width=0.5\linewidth]{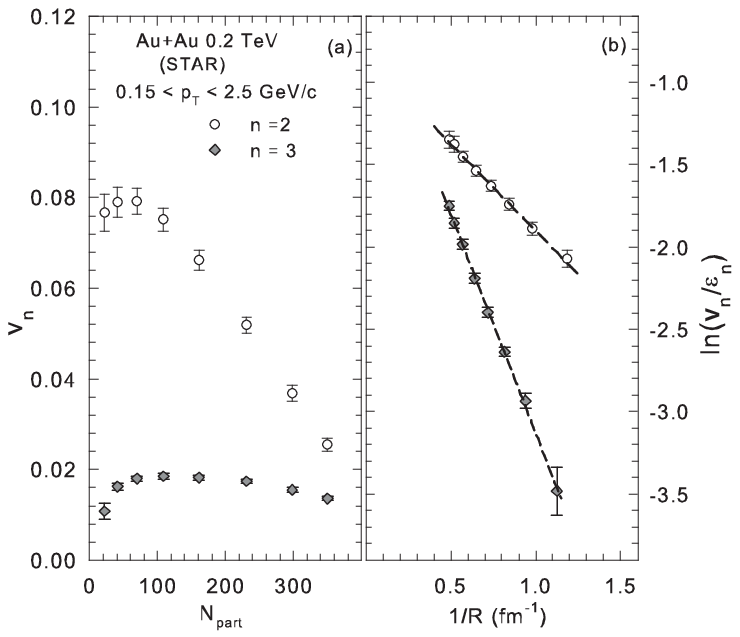} 
\includegraphics[width=0.5\linewidth]{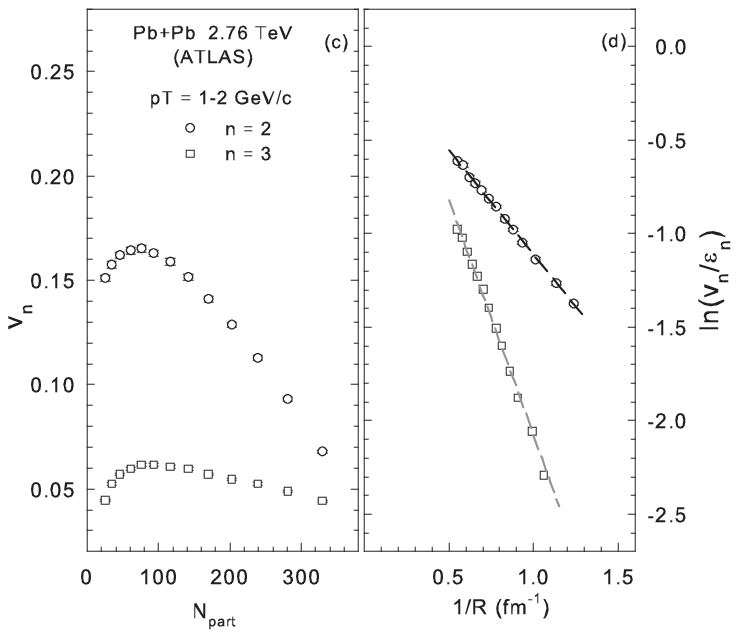}
\end{tabular}
\caption{(a) $p_T$-integrated $v_{2,3}$ vs. N$_\text{part}$ for $0.15 < p_T < 2.5$ GeV/c for Au+Au collisions
at $\sqrt{s_{NN}}= 200$ GeV. The $v_{2,}\text{\{2\}}$ data are taken from Refs. \cite{Adamczyk:2013waa,Agakishiev:2011eq}: 
(b) $\ln(v_n/\varepsilon_n)$ vs. $1/\bar{R}$ for the data shown in (a): 
(c) $v_{2,3}$ vs. N$_\text{part}$ for $p_T = 1-2$ GeV/c for Pb+Pb collisions
at $\sqrt{s_{NN}}= 2.76$ TeV. The latter data are taken from Refs. \cite{ATLAS:2012at,JJia:2011hfa}: 
(d) $\ln(v_n/\varepsilon_n)$ vs. $1/\bar{R}$ for the data shown in (c). The dashed curves in 
(b) and (d) are fits to the data (see text); error bars are statistical only.
}
\label{Fig1}
\end{figure*}
%
In prior work \cite{Lacey:2013is,Lacey:2011ug}, we have validated the acoustic nature of anisotropic flow 
and shown that the strength of the dissipative effects which influence the magnitude of $v_{n}(p_T, \text{cent})$,
can be expressed as a perturbation to the energy-momentum 
tensor $T_{\mu\nu}$ \cite{Staig:2010pn};
\be
\delta T_{\mu\nu} (n,t) = \exp{\left(-\beta' n^2 \right)}  \delta T_{\mu\nu} (0),
 \,\,\, \beta' = \frac{2}{3} \frac{\eta}{s} \frac{1}{\bar{R}^2} \frac{t}{T}, 
\label{eq:3}
\ee 
%
%
%
where $t \propto \bar{R}$ is the expansion time, $T$ is the temperature, $k = n/{\bar R}$ is the wave 
number ({\em i.e.} $2\pi {\bar R} = n\lambda$ for $n\ge 1$), $\bar{R}$ is the initial state transverse size of 
the collision zone and $\delta T_{\mu\nu} (n,0)$ represent the spectrum of initial ($t = 0$) 
perturbations associated with the collision geometry and its density driven fluctuations. 
The latter is encoded in the initial eccentricity ($\varepsilon_n$) moments.
The viscous corrections to ${v_n}/\varepsilon_n$ grow exponentially 
as $n^2$ and $1/\bar{R}$ \cite{Lacey:2013is,Shuryak:2013ke,Lacey:2011ug}; 
\be
\ln\left(\frac{v_n(\text{cent})}{\varepsilon_n(\text{cent})}\right) \propto \frac{-\beta^{''}}{\bar{R}},
\,\,\, \beta'' = \frac{4}{3} \frac{n^2\eta}{Ts},
\label{eq:4}
\ee 
For a given n, Eq.~\ref{eq:4} indicates a characteristic linear dependence of 
$\ln(v_n/\varepsilon_n)$ on $1/\bar{R}$, with slope $\beta'' \propto \eta/s$.
This scaling pattern is borne out in the results of the viscous hydrodynamical calculations \cite{Song:2011hk-a,Song:2011hk} 
shown in Fig.~\ref{Fig1m}. The scaled results, shown for two separate values of $\eta/s$ in Fig.~\ref{Fig1m}(b), 
not only indicate a linear dependence of $\ln(v_n/\varepsilon_n)$ on $1/\bar{R}$, but also a clear sensitivity
of the slopes to $\eta/s$. 
Thus, the validation of this $1/\bar{R}$ scaling for each beam energy, would provide 
a basis for consistent study of the $\sqrt{s_{NN}}$ dependence of the viscous coefficient $\beta''$
\cite{Bonasera:1987zz,Csernai:2006zz,Lacey:2007na}.
Here, we perform such validation tests for the full range of energies available 
at RHIC and the LHC, with an eye towards establishing new constraints for 
the \sqrtsNN~or ($\mu_B, T$) dependence of $\eta/s$. 

	The data employed in our analysis are taken from measurements by the ATLAS and CMS
collaborations for Pb+Pb collisions at $\sqrt{s_{NN}}$ = 2.76 TeV \cite{ATLAS:2012at,JJia:2011hfa,Chatrchyan:2012ta},
as well as measurements by the STAR collaborations for Au+Au collisions spanning the range 
$\sqrt{s_{NN}}= 7.7 - 200$ GeV \cite{Adamczyk:2012ku,Adamczyk:2013waa,Agakishiev:2011eq}. 
The ATLAS and CMS measurements exploit the event plane analysis method and/or the two-particle $\Delta\phi$ 
correlation technique to obtain $v_{n}(p_T,\text{cent})$. To suppress the non-flow correlations, 
a pseudo-rapidity gap ($\Delta\eta_p $) between particles and the event plane, or particle pairs was used. 
The STAR measurements were obtained with several analysis methods for $\sqrt{s_{NN}}= 7.7 - 39$ GeV and 
the Q-cumulant method for $\sqrt{s_{NN}}= 62.4$ and 200 GeV. For purposes of consistency across beam energies, 
we use the data from the event plane analysis method ($v_2\text{\{EP\}}$) for $\sqrt{s_{NN}}= 7.7 - 39$ GeV, and 
the Q-cumulant method ($v_2\text{\{2\}}$) for $\sqrt{s_{NN}}= 62.4$ and 200 GeV. Note that the measurements from 
both analysis methods have been shown to be in good agreement for $\sqrt{s_{NN}}= 7.7 - 39$ GeV \cite{Adamczyk:2012ku}. 
The systematic errors, which are relatively small, are reported in Refs.~\cite{ATLAS:2012at,Chatrchyan:2012ta} 
and \cite{Adamczyk:2012ku,Adamczyk:2013waa,Agakishiev:2011eq} for the respective sets of measurements.

Monte Carlo Glauber (MC-Glauber) simulations were used to compute the number of participants N$_\text{part}(\text{cent})$, 
participant eccentricity $\varepsilon_n({\text{cent}})$ [with weight $\omega(\mathbf{r_{\perp}}) = \mathbf{r_{\perp}\!}^n$]
(and $\varepsilon_n\{2\}({\text{cent}})$) and $\bar{R}({\text{cent}})$ from the two-dimensional profile of the density of 
sources in the transverse  plane $\rho_s(\mathbf{r_{\perp}})$ \cite{Lacey:2010hw}; 
${1}/{\bar{R}}~=~\sqrt{\left({1}/{\sigma_x^2}+{1}/{\sigma_y^2}\right)}$, 
where $\sigma_x$ and $\sigma_y$ are the respective root-mean-square widths of the density distributions. 
The initial state geometric quantities so obtained, are in excellent agreement with the values 
reported for Pb+Pb collisions at $\sqrt{s_{NN}}$ = 2.76 TeV \cite{Chatrchyan:2012ta} 
and Au+Au collisions for the range $\sqrt{s_{NN}}= 7.7 - 200$ GeV \cite{Adamczyk:2012ku,Agakishiev:2011eq}.
A centrality independent systematic uncertainty estimate of 2-3\% was obtained for $\bar{R}$ and $\varepsilon$ respectively,
via variations of the MC-Glauber model parameters.  
We use the values of $\bar{R}$ and $\varepsilon$ in concert with the RHIC and LHC data sets to perform validation 
tests for $1/\bar{R}$ scaling over the centrality selections of 5-70\%, for each of the available beam energies.
%
%
\begin{figure*}[t]
\includegraphics[width=1.0\linewidth]{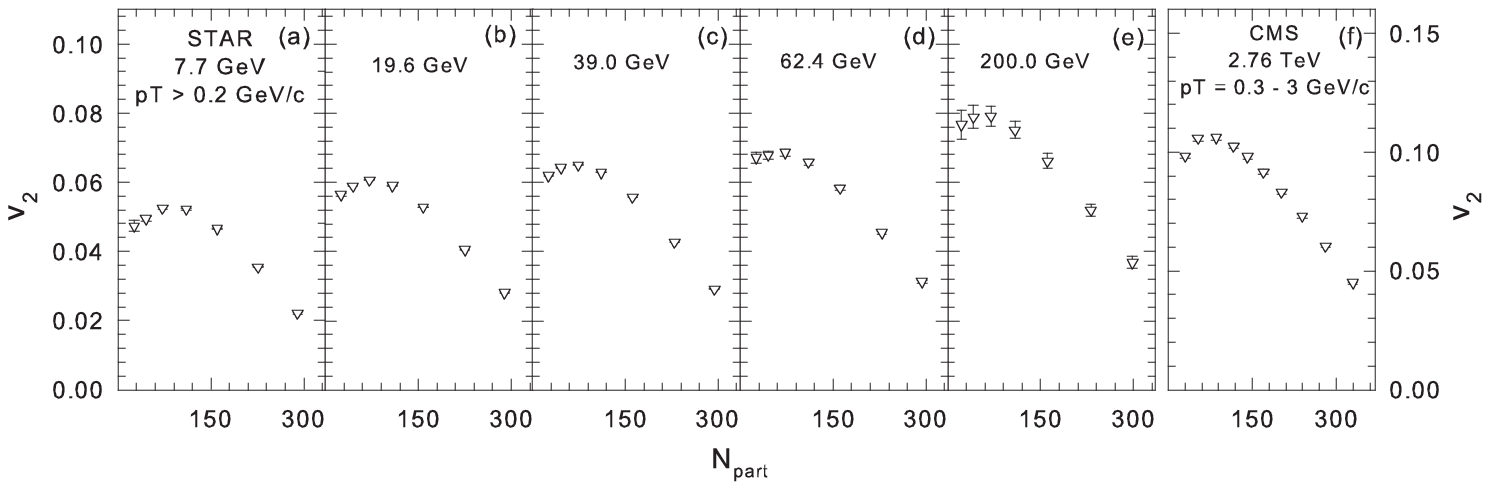}
\vskip 0.8cm
\includegraphics[width=1.0\linewidth]{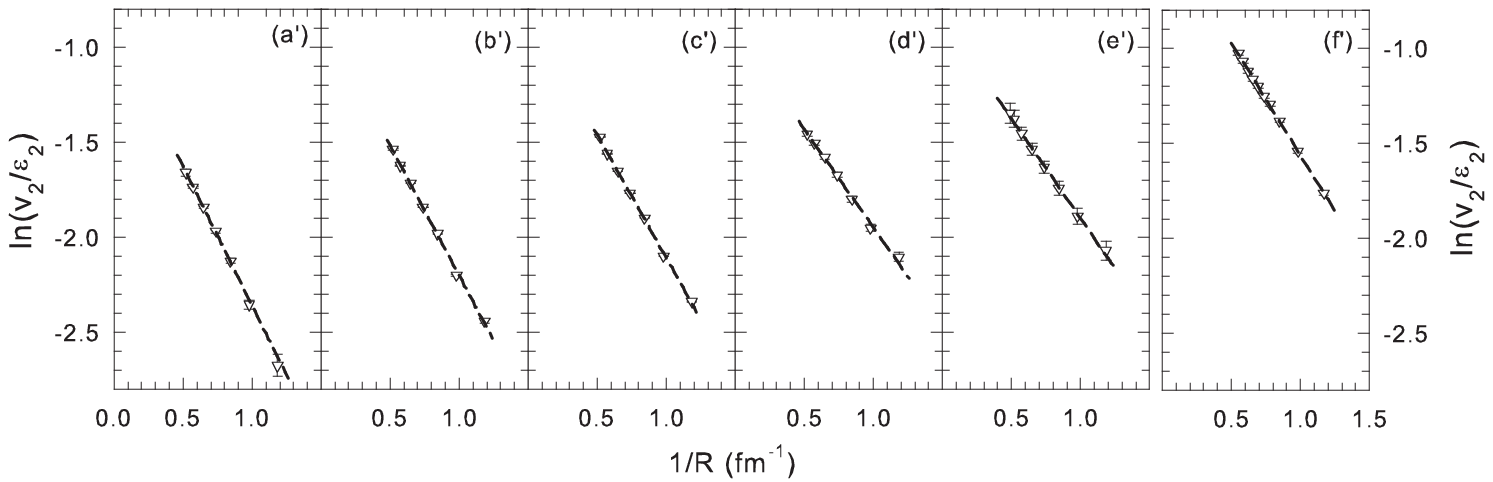}
\caption{((a)-(e)) $p_T$-integrated $v_{2}$ vs. N$_\text{part}$ for $p_T \agt 0.2$ GeV/c for Au+Au collisions
for several values of $\sqrt{s_{NN}}$ as indicated. The data are taken from Refs. \cite{Adamczyk:2012ku,Agakishiev:2011eq}: 
(f) $v_{2}$ vs. N$_\text{part}$ for $p_T = 0.3 - 3$ GeV/c for Pb+Pb collisions
at $\sqrt{s_{NN}}=2.76$ TeV. These data are taken from Ref. \cite{Chatrchyan:2012ta}: 
(($\text{a}')-(\text{f}'$)) $\ln(v_2/\varepsilon_2)$ vs. $1/\bar{R}$ for the data shown in (a)-(f). 
The dashed curves are linear fits to the data; error bars are statistical only.
}
\label{Fig2}
\end{figure*}

Figures~\ref{Fig1}(a) and (c) show representative plots of $v_{2,3}$ vs. N$_\text{part}$ for 
Au+Au and Pb+Pb collisions respectively. They show that $v_{2,3}$ increases 
from central ($\text{N}_{\text{part}} \sim 340$) to mid-central ($\text{N}_{\text{part}} \sim 120$) collisions 
as would be expected from an increase in $\varepsilon_{2,3}$ over the same $\text{N}_{\text{part}}$ range. 
For $\text{N}_{\text{part}} \alt 120$ however, the decreasing trend of $v_{2,3}$ contrasts with the known increasing 
trends for $\varepsilon_{2,3}$, suggesting that the viscous effects due to the smaller systems produced in peripheral
collisions, serve to suppress $v_{2,3}$. This is confirmed by the symbols and dashed curves in Figs.~\ref{Fig1}(b) 
and (d) which validate the expected linear dependence of $\ln(v_n/\varepsilon_n)$ on $1/\bar{R}$  (cf. Eq.~\ref{eq:4}) 
for the data shown in Figs.~\ref{Fig1}(a) and (c). Note that the slopes for $n=3$ are more than a factor of 
two larger than those for $n=2$ as expected (cf. Eq.~\ref{eq:4}). A similar dependence was observed 
for other $p_T$ selections.

\begin{figure}[t]
\includegraphics[width=1.0\linewidth]{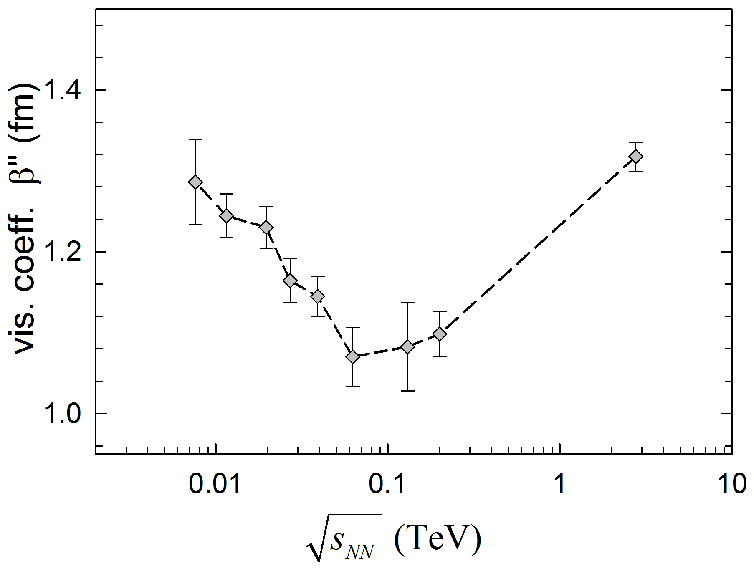}
\vskip 0.10cm
\caption{Viscous coefficient $\beta''$ vs. $\sqrt{s_{NN}}$, extracted from linear 
fits to $\ln(v_2/\varepsilon_2)$ vs. $1/\bar{R}$; error bars are statistical only. 
The dashed curve is drawn to guide the eye.
}
\label{Fig3}
\end{figure}

Validation tests for this $1/\bar{R}$ scaling of $v_2$ were carried out for the full range of available 
beam energies as illustrated in Fig.~\ref{Fig2}. Figs.~\ref{Fig2}(a)-(f) show $p_T$-integrated $v_{2}$ vs. N$_\text{part}$ 
for a representative set of these collision energies as indicated; they show the same characteristic 
pattern observed for $v_2$ in Figs.~\ref{Fig1}(a) and (c). That is, the increase in $v_2$ from central 
to mid-central collisions, followed by a decrease for peripheral 
collisions, persists across the full range of collision energies.  
We interpret this as an indication that the transverse size of the collision zone plays 
a similar mechanistic role in viscous damping across the full range of beam energies studied. 
This is further confirmed in Figs.~\ref{Fig2}($\text{a}')-(\text{f}'$) which show the expected 
linear dependence of $\ln(v_n/\varepsilon_n)$ vs. $1/\bar{R}$, for the data 
shown in Figs.~\ref{Fig2}(a)-(f). A similar dependence was observed for the other collision 
energies ($\sqrt{s_{NN}}=$~11.5, 27 and 130 GeV) not shown in Fig.~\ref{Fig2}. This pervasive pattern of scaling provides 
the basis for a consistent method of extraction of the viscous coefficient $\beta^{''} \propto \eta/s$,
via linear fits to the scaled data for each beam energy. 
The dashed curves in Figs.~\ref{Fig2}($\text{a}')-(\text{f}'$) show representative examples 
of such fits. The $\beta^{''}$ values, with statistical errors obtained from these fits, 
are summarized in Fig.~\ref{Fig3}.

Figure~\ref{Fig3} indicates only a mild variation in the magnitude of $\beta^{''}$ for the 
broad span of collision energies studied (note the factor of $\sim 360$ increase from RHIC BES to
LHC). This variation is compatible with the 
observation that $v_2$ measurements, obtained at different $\sqrt{s_{NN}}$, show similar  
magnitudes. That is, a larger variation of these coefficients would necessitate a much larger 
variation in the $v_2$ values obtained at different values of $\sqrt{s_{NN}}$, because of viscous damping.
A more striking feature of Fig.~\ref{Fig3} is the $\sqrt{s_{NN}}$ dependence of $\beta^{''}$.
It shows that $\beta^{''}$ decreases as $\sqrt{s_{NN}}$ increases from $7.7$~GeV to 
approximately 62.4~GeV, followed by a relatively slow increase from $\sqrt{s_{NN}}= 62.4$~GeV - 2.76~TeV. 
Here, it should be emphasized that the error bars for the extractions made at 62.4, 130 and 200 GeV, as well as a  
lack of measurements between 39 and 62.4 GeV, do not allow a definitive estimate of the 
actual location of this apparent minimum. Nonetheless, we interpret this trend 
as an indication for the change in $\left<\eta/s\right>$ which results from the difference in the decay trajectories 
sampled [in the $(T,\mu_B)$ plane] at each collision energy \cite{Csernai:2006zz,Lacey:2007na}. A similar qualitative 
pattern of viscous damping has been recently obtained in transport calculations~\cite{Ozvenchuk:2012kh}, as well as 
to reconcile the similarity between charged hadron $v_2(p_T)$ measurements 
obtained for $\sqrt{s_{NN}} > 62.4$~GeV \cite{Plumari:2013bga}.

The characteristic $\sqrt{s_{NN}}$ dependence of $\beta^{''}$ shown in Fig.~\ref{Fig3}, also bears a 
striking resemblance to the $T$ and $\mu_B$ dependence of $\eta/s$ for atomic and molecular 
substances, which show ${\eta}/{s}$ minima with a cusp at the critical end 
point $(T_{\text{cep}},\mu^{{\text{cep}}}_B)$  \cite{Csernai:2006zz,Lacey:2007na}.
Thus, the observed trend of the $\sqrt{s_{NN}}$ dependence of $\beta^{''}$ could also be an indication 
for decay trajectories which lie close to the CEP. Further detailed extractions of $\beta^{''}$, with 
reduced error bars, are however required to pin point the apparent minimum and to further 
confirm its relationship to a possible CEP.

%
%
%


In summary, we have presented a detailed phenomenological study of viscous damping of the flow 
harmonics $v_{2,3}$ for Pb+Pb collisions at $\sqrt{s_{NN}}= 2.76$ TeV, and 
for Au+Au collisions spanning the range $\sqrt{s_{NN}}= 7.7 - 200$~GeV. Our study shows 
that this damping can be understood to be a consequence of the acoustic nature of 
anisotropic flow. That is, it validates the characteristic signature expected for 
the system size dependence of viscous damping [at each collision energy] 
inferred from the dispersion relation for sound propagation in the 
matter produced in the collisions. The extracted viscous coefficients, which 
encode the magnitude of the ratio of shear viscosity to 
entropy density $\eta/s$, are observed to decrease to an apparent minimum 
as the collision energy is increased from $\sqrt{s_{NN}}= 7.7$ to 62.4~GeV, albeit with a sizeable error; thereafter, it 
shows a slow increase with $\sqrt{s_{NN}}$. This pattern of viscous damping provides a first indication 
for the variation of $\eta/s$ in the temperature-baryon chemical potential ($T, \mu_B$) plane. 
It also bears a striking resemblance to the observations for atomic and molecular substances, 
which show ${\eta}/{s}$ minima with a cusp at the critical end point $(T_{\text{cep}},\mu^{{\text{cep}}}_B)$. 
Further detailed studies, with improved errors and other harmonics, are required to make a more precise 
mapping of viscous damping in the ($T, \mu_B$)-plane, as well as to confirm if the observed pattern 
for $\beta''(\sqrt{s_{NN}})$ reflects decay trajectories close to the critical end point 
in the phase diagram for nuclear matter.

{\bf Acknowledgments:}
This research is supported by the US DOE under contract DE-FG02-87ER40331.A008 and by the NSF under award number PHY-
1019387
 


%
\bibliography{viscous_coefficient} 
\end{document}